# Quantum effects in graphitic materials: Colossal magnetoresistance, Andreev reflections, Little-Parks effect, ferromagnetism, and granular superconductivity

Nadina Gheorghiu[1], Charles R. Ebbing[2], Benjamin T. Pierce[3], and Timothy J. Haugan[3]

[1]*UES Inc., Dayton, OH 45432*

[2] *University of Dayton Research Institute, Dayton, OH 45469*

[3]*The Air Force Research Laboratory, Wright-Patterson Air Force Base, OH 45433*

Nadina.Gheorghiu@yahoo.com

**Abstract**. Unlike the more common local conductance spectroscopy, nonlocal conductance can differentiate between nontopological zero-energy modes localized around inhomogeneities, and true Majorana edge modes in the topological phase. In particular, negative nonlocal conductance is dominated by the crossed Andreev reflection. Fundamentally, the effect reflects the system's topology. In graphene, the Andreev reflection and the inter-band Klein tunneling couple electron-like and hole-like states through the action of either a superconducting (SC) pair potential or an electrostatic potential. We are here probing quantum phenomena in modified graphitic samples. Four-point contact transport measurements at cryogenic to room temperatures were conducted using a Quantum Design Physical Property Measurement System. The observed negative nonlocal differential conductance $G_{diff}$ probes the Andreev reflection at the walls of the SC grains coupled by Josephson effect through the semiconducting matrix. In addition, $G_{diff}$ shows the butterfly shape that is characteristic to resistive random-access memory devices. In a magnetic field, the Andreev reflection counters the effect of the otherwise lowered conduction. At low temperatures, the magnetoresistance shows irreversible yet strong colossal oscillations that are known to be quantum in nature. In addition, we have found evidence for seemingly granular SC as well as ferromagnetism. Moreover, the Little-Parks effect is revealed in both the classical small-amplitude and the phase-slip driven large-amplitude oscillations in the magnetoresistance. Thus, graphitic materials show potential for quantum electronics applications, including rectification and topological states.

## 1. Introduction

While carbon(C)-based materials are known for many practical properties, such as light-weight and high strength, their inner nature is governed by the laws of quantum physics manifested as magnetism [1], conveniently used for spintronics [2], or as unconventional superconductivity (SC) [3,4] that we have previously explored [5]. In this paper we present evidence for several quantum phenomena occurring in graphitic materials: nonlinear electronic transport resulting in negative differential resistance, colossal magnetoresistance, Andreev reflections at the boundaries between SC grains and the surrounding semiconducting matrix, and ferromagnetism (FM), and field-dependent magnetoresistance oscillations due to the Little-Parks effect.

## 2. Experimental details

The materials are PAN-based (polyacrylonitrile ($CH_2$-CH-CN)$_n$) T300 C fibers (Cytec), highly oriented pyrolytic graphite (particle size 10 μm), and graphite foil (Graphtek). In order to increase the density of states to levels where possible SC behaviour can be observed, selected samples were oxygen(O)-ion implanted (Cutting Edge Ions) at 2D concentrations $7.07 \times 10^{12}$, $5.66 \times 10^{15}$, and $2.24 \times 10^{16}$ ions/cm$^2$.

For an implantation energy 70 keV, the implantation depth was 125 nm for graphite and 80 nm for diamond-like C films, respectively. The four-wire Van Der Pauw technique was employed for resistivity measurements. Scanning electron microscopy pictures and the elemental analysis for C fiber samples are shown in Fig. 1. The quality of the electrical contacts was optically checked using an Olympus BX51 microscope. The current-to-voltage gap ratio was close to the required factor of four [6]. Magneto-transport and magnetization measurements were carried out in the 1.9 K - 300 K temperature range and for magnetic fields of induction $B$ up to 9 T using a Physical Properties Measurement System (PPMS) model 6500 (Quantum Design).

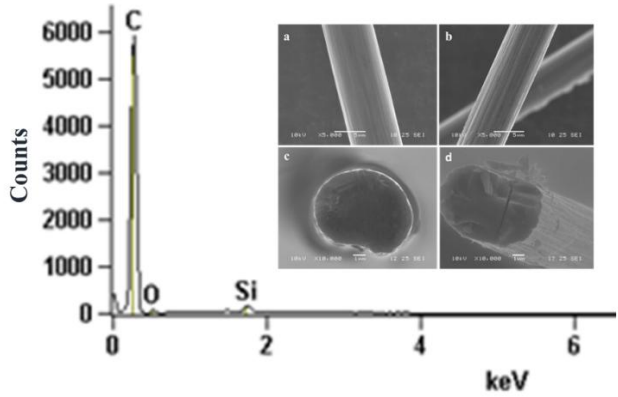

**Figure 1** SEM pictures showing the effect of O-ion implantation ($2.24 \times 10^{16}$ ions/cm$^2$) on the C fiber (b, d) vs. to the raw C fiber (a, c), both along and across the fibers (upper images and lower images, respectively). Also shown are the C and O peaks on the EDS scan. The Si peak is due to the placement of the C fiber on a Si substrate during the implantation procedure.

## 3. Results and Discussions
*Electrical resistivity and nonlocal differential conductance without magnetic field*

The temperature-dependent resistivity $\rho(T)$ for two C fibers (#1 and #2) O-implanted at $2.24 \times 10^{16}$ ions/cm$^2$ is shown in Fig. 2. At $T \sim 250$ K, there is either an insulator-metal-insulator transition or a metal-insulator-metal transition for a sourced current $I = 20$ μA or $I = 1$ μA, respectively. These transitions reflect the $T$-dependence balance of charge carrier densities and their mobilities. For the larger $I$, $\rho$ has a metallic behaviour below $T \sim 25$ K [7]. For the smaller $I$, $\rho$ for one of the fibers becomes negative for temperatures below ~150 K. Negative resistance has been also related to the cooling of the Fermi fluid by Joule-Thomson effect [8]. The observation of negative resistance (i.e., nonlocal negative voltage) is a signature of highly viscous flow of electrons or, equivalently, zero electrical resistance in the SC state [9]. Moreover, the BCS mean-field critical temperature for Cooper-pair instability in graphene is $T_c \sim 150$ K [10].

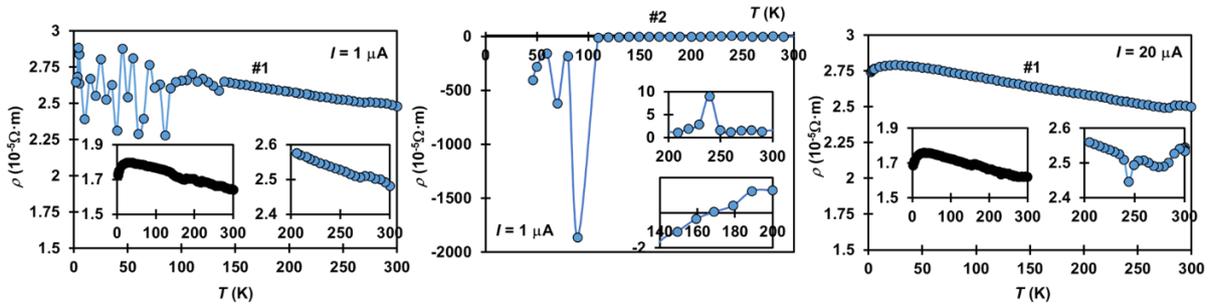

**Figure 2** $\rho(T)$ for two C fibers O-implanted at $2.24 \times 10^{16}$ ions/cm$^2$ (in blue) and raw fiber (in black).

Below 150 K, giant fluctuations in the nonlinear differential conductance $G_{diff}$ = d$I$/d$V$ and the Chua's butterfly are observed at $T$ = 130 K (Fig. 3). For $V < 0$, $G_{diff} > 300$ mS or $R_{diff} < 4$ Ω or $\rho_{diff} < 8$ μΩ·cm. The Chua nonlinear phenomenon can be also seen as a subcritical Hopf bifurcation in which new attractors are being born at the pitchfork point [11]. The inset for V > 0 shows that $G_{diff}$ has the shape of the nonlinear $I(V)$ in the Chua circuit. I.e., the charge oscillates as a third-order polynomial of the magnetic flux as in a cubic memristor during a supercritical Andronov–Hopf bifurcation. A [12]. The butterfly wing span is $\Delta V \cong 0.5$ mV, which would correspond to the critical temperature for twisted bilayer graphene, $T_c \cong 1.8$ K. Interestingly, the minimum $G_{diff}$ (< 0) corresponding to $V \cong 60$ mV would give a BCS critical temperature $T_c = 2\Delta/3.53k_B \cong 250$ K, exactly at the insulator-metal transition. The subcritical and supercritical bifurcations meet at the tricritical point and the magnetic field of the local spins acts as a chemical potential. The oxide layer at the surface of the C fiber might host both hard and soft SC fluctuations, i.e., the system might be a multigap SC where the larger gap SC phase takes over the smaller gap SC phases. The evolution of the attractor – which is a manifold of equilibrium states – is actually the evolution of the amplitude (Higgs mode) and phase for the SC order parameter.

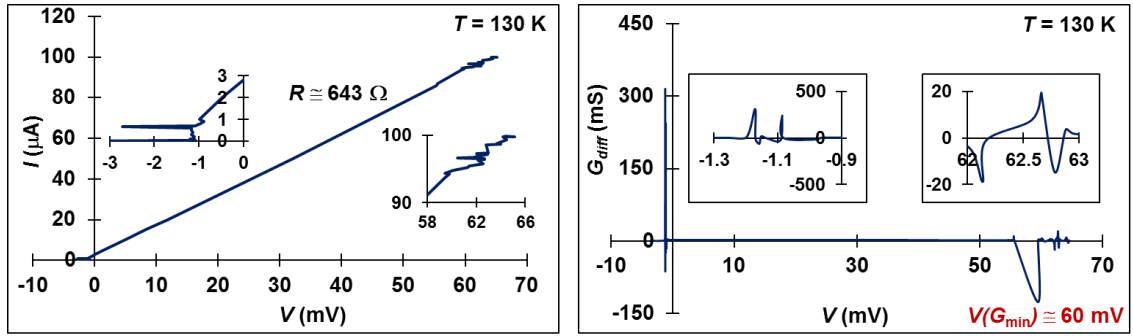

**Figure 3** $I(V)$ and the corresponding $G_{diff}(V)$ data for O-implanted C fiber (2.24 × 10$^{16}$ ions/cm$^2$).

3.1. *Transport measurements in magnetic field*
3.2.2 *Integer and fractional universal (quantum) conductance peaks and butterfly-shaped resistance-current curves*

As known, graphite has antiferromagnetic (AFM) correlations between unlike sublattices (ABAB...) and FM correlations between like sublattices (AAA... or BBB...). A magnetic field can turn a FM into an AFM at the tricritical point, where the FM and AFM merged into the system paramagnetic (PM) phase as the temperature is increased. Under a static magnetic field, the differential resistance $R_{diff} = dV/dI$ and not $V(I)$ shows the butterfly shape (Figure 4). Significantly, we observe several extrema in the differential conductance $G_{diff}$. A maximum peak on the electron side ($V > 0$) is at the perfect conductance quantum $G_0 = 2e^2/h \cong 77$ μS. On the hole side ($V < 0$), one maximum peak is at $G_0/4$ and a minimum peak is at $\cong -G_0/2$. As we will see later, the peaks in $G_{diff}$ are signatures for the Andreev states. Significantly, the $G_{diff} < 0$ signature of crossed Andreev reflection suggests the presence of SC-metal-SC-metal-SC interfaces, i.e., acting also as Josephson junctions.

3.2.2 *Colossal magnetoresistance*

Low-temperature magnetoresistance $MR(\%) = 100 \times [R(B) - R(B = 0)]/R(B = 0)$ for an O-implanted (2.24 × 10$^{16}$ ions/cm$^2$) C fiber is shown in Figure 5. While $MR$ takes colossal values for both small (50 nA) and large (50 μA) currents, the electronic charge transport is significantly less viscous for the former ($MR < 0$) than for the latter ($MR > 0$). Viscosity generates vorticity, arguable playing the same role as the zero-electrical resistance does for SC [9]. A colossal *MR* material can be used a substrate for another high-temperature SC material like YBCO or BSCCO material for significant flux injection and thus improved pinning.

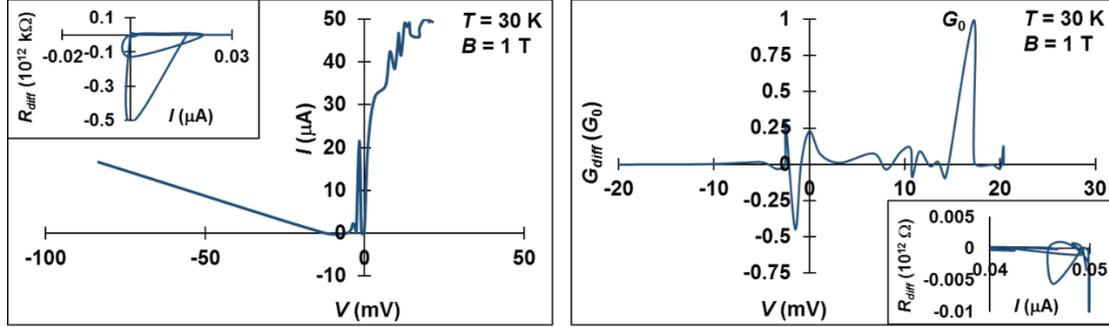

**Figure 4** *I(V)* and the corresponding $G_{diff}(V)$ data for O-implanted C fiber ($2.24 \times 10^{16}$ ions/cm$^2$) in transverse magnetic field. Inset: Butterfly-shaped differential resistance vs. current.

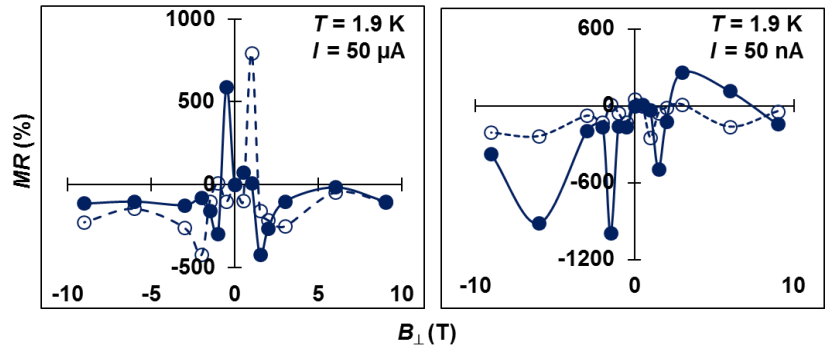

**Figure 5** Low-temperature magnetoresistance for an O-implanted ($2.24 \times 10^{16}$ ions/cm$^2$) C fiber.

*3.2.2 Andreev oscillations*

We have also observed Andreev oscillations in the *MR* of highly oriented pyrolytic graphite (HOPG) cylindrical, disc-like samples of radius $R = 3$ mm and thickness 1 mm (Figure 6). The magnetic field was applied along the disc's axis and a 20 μA current was passed through the Van der Pauw wire arrangement. While this kind of quantum resonances that are attributed to Andreev reflections between SC and semiconducting regions have been observed before [13], the reported period was significantly larger (0.1 T to 0.4 T). Here, the period for Andreev oscillations is $\Delta B = 2$ mT. We can estimate the area for the SQUID-type oscillations of two SC pathways separated by a normal/insulating region as $S = \Phi_0/\Delta B \cong 1.0 \times 10^{-12}$ m$^2$, where $\Phi_0 \cong 2.07 \times 10^{-15}$ T·m$^2$ is the magnetic flux quanta. Given the sample's symmetry, this would give a radius $R \cong 560$ nm for the circular paths. In contrast, the large-amplitude period of ~1 - 1.5 T observed for the wire (C fiber) sample (Figure 5) gives a radius of only 21 - 26 nm that is close to the mean distance along relatively straight portions of the PAN-fibers' graphitic network, $l \cong 10$ nm [14]. Indeed, the slightly wave-like pattern of the C fiber can accommodate fluxoids that are pinned at grain boundaries and extend over four nearly-straight graphitic domains. While small-amplitude oscillations in the resistance are due to the Little-Parks effect [15,16], the large-amplitude oscillations are attributed to the field-driven modulation of barrier heights for phase slips (i.e., thermal fluctuations) [17]. Half-fluxoids ($R \sim 15 - 18$ nm) would indicate a chiral p-wave triplet SC [18] that might also coexist with FM [19] (see magnetization loops below).

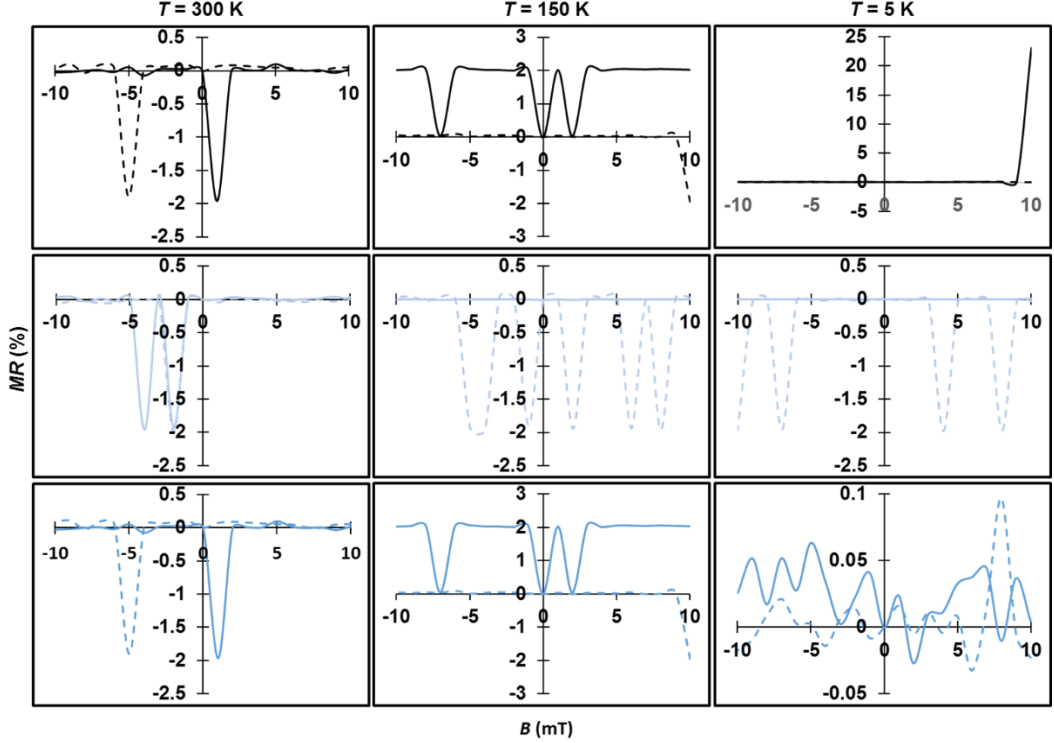

**Figure 6** Andreev oscillations in the magnetoresistance of HOPG cylindrical samples O-implanted at $7.07 \times 10^{12}$ ions/cm² (light blue), $5.66 \times 10^{15}$ ions/cm² (blue), and raw sample (in black) at three different temperatures.

*3.3 Magnetization measurements*
*3.3.1 Zero-field cooled (ZFC) and field-cooled (FC) measurements*

Temperature-dependent ZFC-FC magnetization data for O-implanted graphite foil square samples having dimensions 2 mm × 2 mm × 1 mm are shown in Figure 7a. While fluctuations are observed up to $T \sim 150$ K, the irreversibility temperature is $T_{irrev} \cong 50 - 60$ K. Our results are close to the BCS mean-field calculations that found the critical temperature $T_c$ for graphite at 60 K [20] and for graphene at 150 K [21]. Interestingly, the $G_0$ peak at $V \cong 17.3$ mV (Figure 4) gives a BCS critical temperature $T_c = |e|V/3.53k_B \cong 57$ K, where $e \cong -1.6 \times 10^{-19}$ C is the electron's charge and $k_B \cong 1.38 \times 10^{-23}$ J/K is Boltzmann's constant.

An unusual feature is the upward turn of the magnetization at low temperatures that was interpreted as a reentrant phenomenon [22]. The SC-like transition below $T_c \sim 50$ K was also found for diamond-like C films (Figure 7b). Interestingly, the O-implanted films show metastability between the SC and the normal state that is diminished by a transverse magnetic field (of induction $B$). Negative resistance (hole conduction) dominates from ~200 K to ~50 K. I.e., it is possible that short-range SC fluctuations starting at $T \sim 200$ K eventually grow into long-range SC fluctuations that become critical at $T \sim 50$ K. In addition, the insulator-metal transition observed at $T \sim 250$ K is seemingly not affected by $B$. Fullerene $C_{60}$ buckyball clusters could be formed as a result of O-ion implantation. The metal-insulator transition at $T \sim 250$ K could be a structural phase transition. In the solid state, the $C_{60}$ molecules assume a simple cube unit cell that exhibits a phase transition (upon increasing the temperature) to a face-centered-cubic structure upon heating around a transition temperature ~250 K [23].

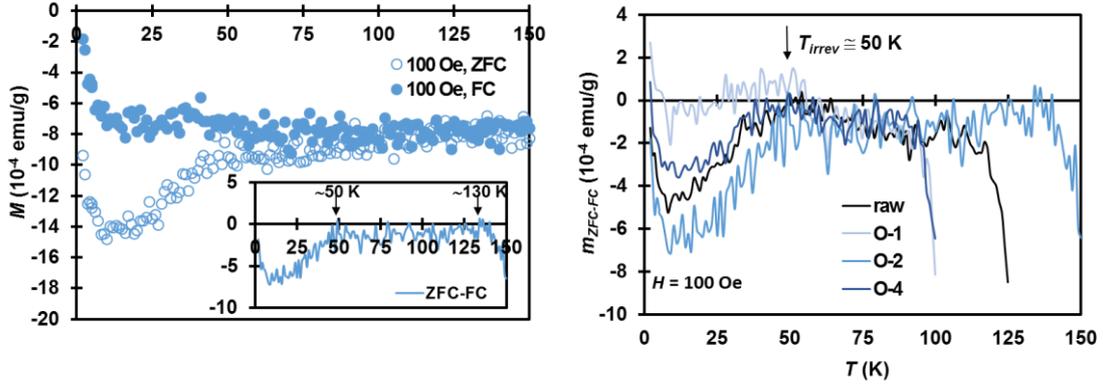

**Figure 7a** Zero-field cooled (ZFC) and field-cooled (FC) temperature dependent magnetization data for graphite foil samples O-implanted at 2D concentrations 7.07 × $10^{12}$ ions/cm$^2$ (O-1), 5.66 × $10^{15}$ ions/cm$^2$ (O-2), and 2.24 x $10^{16}$ ions/cm$^2$ (O-3). Data for a raw sample is also shown.

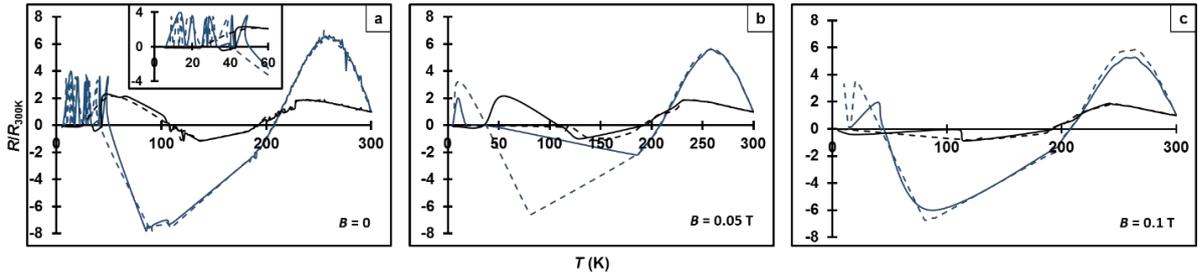

**Figure 7b** Temperature dependent resistance data (relative to the resistance at 300 K) for diamond-like C films O-implanted at 2.24 x $10^{16}$ ions/cm$^2$ (dark blue) vs. raw samples (black), without (a) and with a transverse magnetic field $B_\perp$ = 0.05 T (b) and $B_\perp$ = 0.1 T (c), respectively. Dash/full line for data acquired during decreasing/increasing the temperature.

*3.3.2 Magnetization loops*

Magnetic-field dependent magnetization loops at $T$ = 300 K for the described before graphite foil samples (Figure 8) show that the SC-like diamagnetism coexists with FM. Except for the prolonged tail, these high-temperature magnetization loops are clearly similar to the ones found for hard SCs. In addition, these loops also show the granularity kink characteristic to granular SCs such as the ones found in [24]. Low-field loops (Figure 9a) show the shielding effect due to the pseudo-Meissner effect [25]. The M(H) loops for HOPG and graphite samples are similar (Figure 9b). The magnetic coherence length $\xi_m = (\Phi_0/2\pi B_{c2})^{1/2}$ is in the 100 – 200 nm range.

## 4. Conclusions

Graphitic materials are extremely rich in quantum phenomena. In this work, we have presented experimental evidence for Andreev reflections, colossal magnetoresistance, Little-Parks effect, and granular superconducting-like behaviour coexisting with ferromagnetism. Future work will explore in more detail these and other related phenomena.

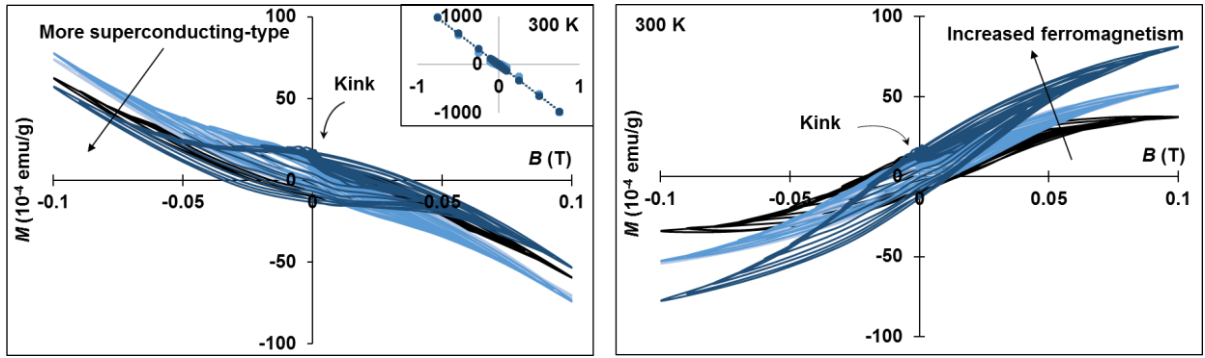

Figure 8 Nonlinear and irreversible $M(B)$ loops at $T = 300$ K for O-implanted graphite foil samples: raw (in black) and O-implanted (in blue, description in Figure 7) before (left) and after (right) the subtraction of their own diamagnetic background (inset).

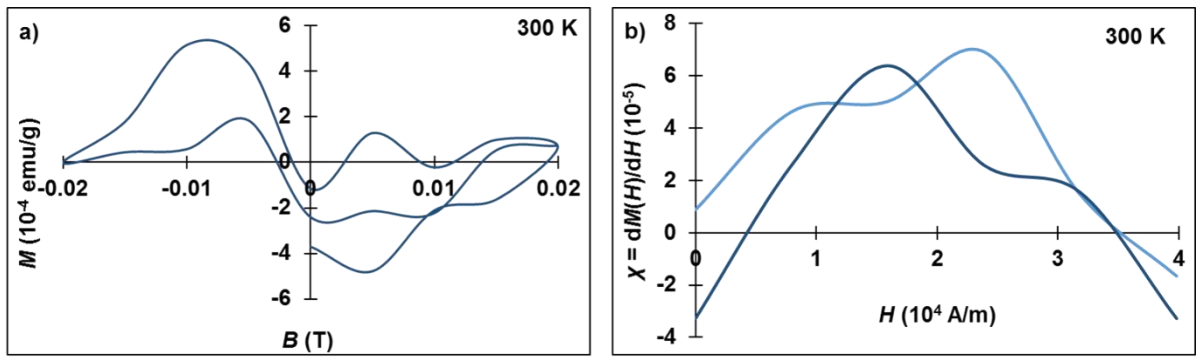

**Figure 9** a) Superconducting-type magnetization loop for a graphite foil sample (O-ion implantation dose $5.64 \times 10^{15}$ ions/cm$^2$). b) Field-dependence of the magnetic susceptibility $\chi$ for an O-implanted sample at $2.24 \times 10^{16}$ ions/cm$^2$ (in dark blue) and for an O-implanted (at $5.64 \times 10^{15}$ ions/cm$^2$) graphite foil sample (in blue).

**Acknowledgments**
This work was supported by The Air Force Office of Scientific Research (AFOSR) for the LRIR #14RQ08COR & LRIR #18RQCOR100 and the Aerospace Systems Directorate (AFRL/RQ).
We acknowledge J.P. Murphy for the cryogenics, and Dr. T.J. Bullard for suggestions.
Dr. G.Y. Panasyuk is acknowledged for his experienced view on condensed matter physics topics and for continuous support and inspiration to this paper's first author.